\documentclass{optica-article}

\journal{opticajournal} 

\articletype{Research Article}

\usepackage{lineno}

\usepackage{icomma}
\usepackage{bm}     
\usepackage{amsmath}    
\usepackage{mathtools}
\usepackage{derivative}

\usepackage{siunitx}    

\usepackage{graphicx}   
\usepackage{wrapfig}    
\usepackage{float}      
\usepackage{url}
\usepackage[export]{adjustbox}

\usepackage{caption}
\usepackage{subcaption}
\usepackage{etoolbox}

\appto
\usepackage[margin=1cm]{caption}
\usepackage{siunitx}
\usepackage{array}
\usepackage{multirow}
\usepackage{hyperref}

\newcommand\animagea{\adjustbox{valign=m,vspace=1pt}{\includegraphics[width=0.28\linewidth]{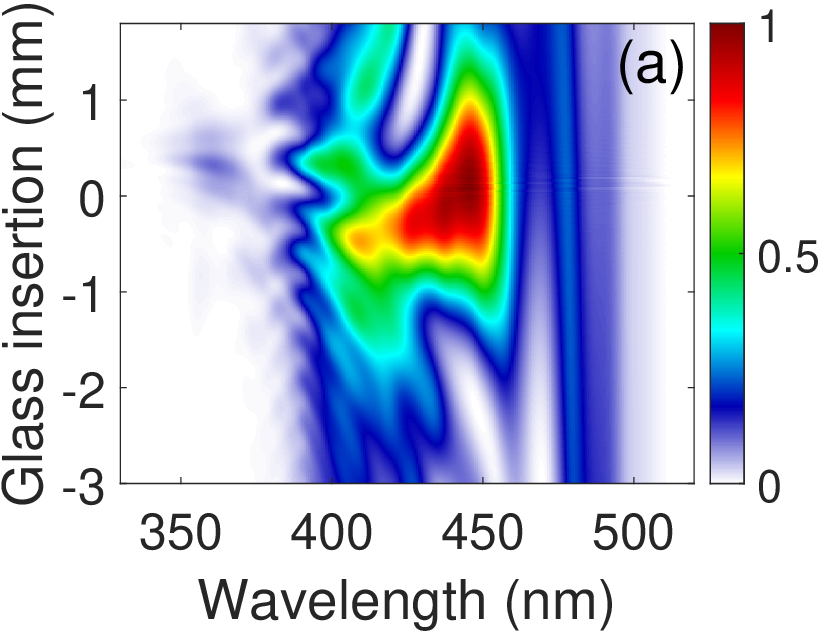}}}
\newcommand\animageb{\adjustbox{valign=m,vspace=1pt}{\includegraphics[width=0.28\linewidth]{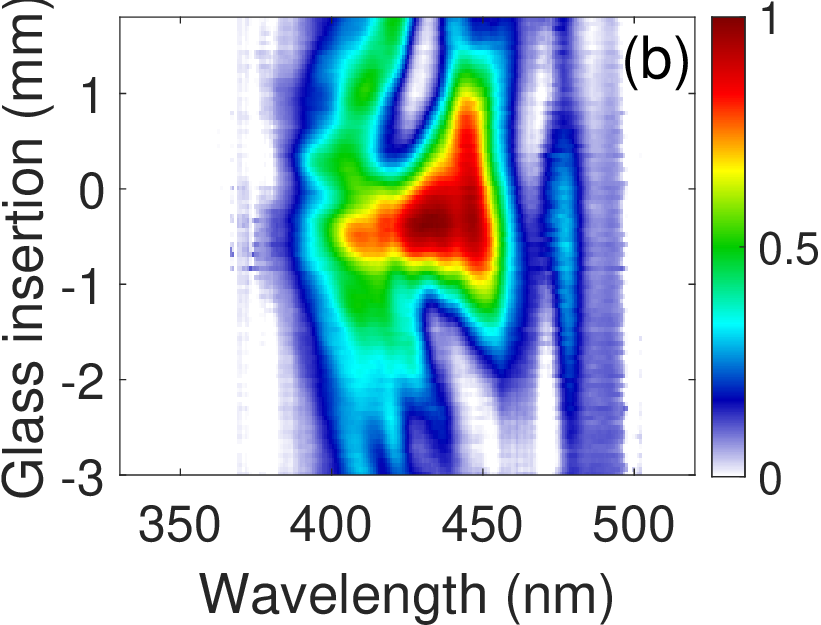}}}
\newcommand\animagec{\adjustbox{valign=m,vspace=1pt}{\includegraphics[width=0.28\linewidth]{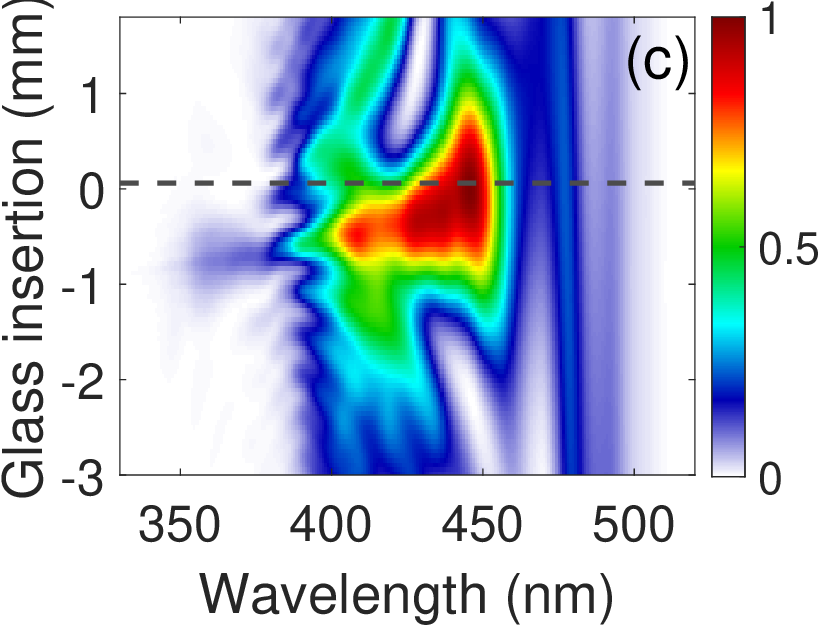}}}
\newcommand\animaged{\adjustbox{valign=m,vspace=1pt}{\includegraphics[width=0.28\linewidth]{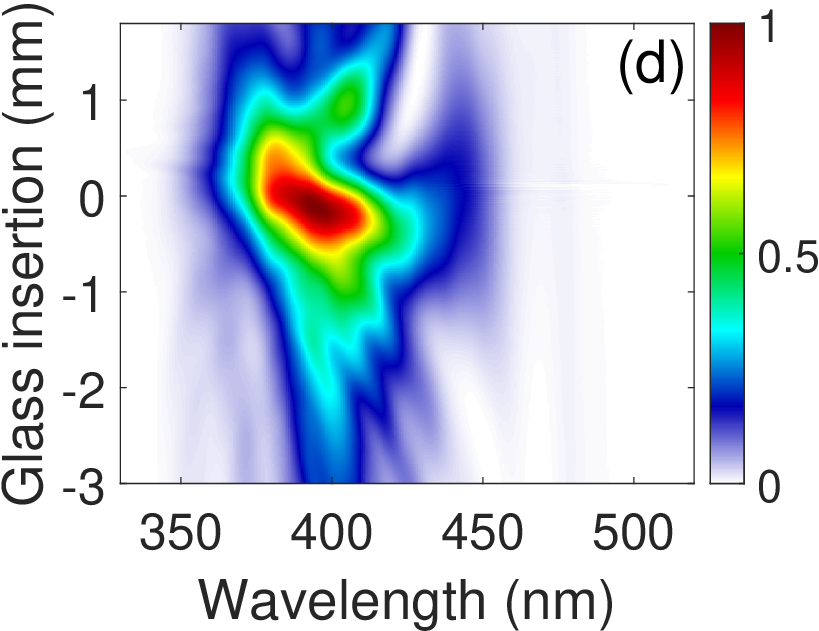}}}
\newcommand\animagee{\adjustbox{valign=m,vspace=1pt}{\includegraphics[width=0.28\linewidth]{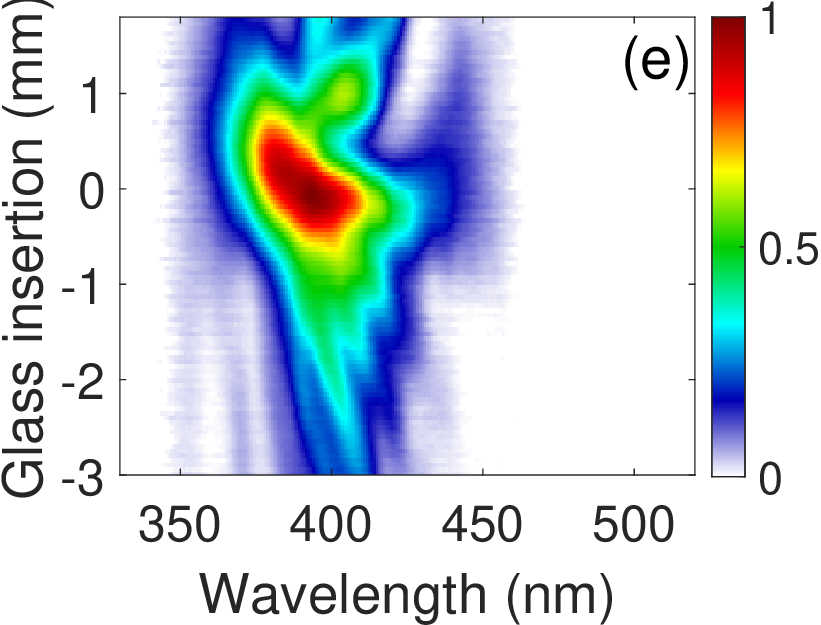}}}
\newcommand\animagef{\adjustbox{valign=m,vspace=1pt}{\includegraphics[width=0.28\linewidth]{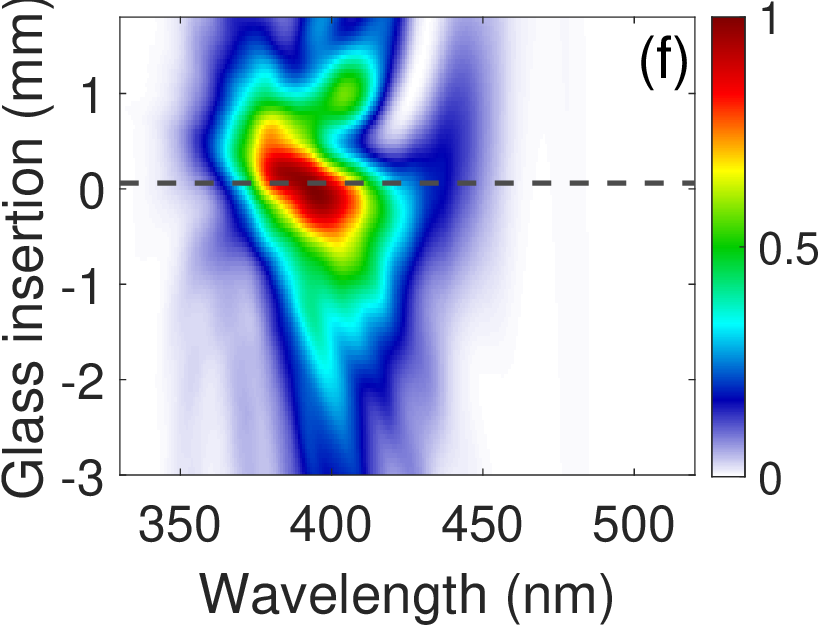}}}
\newcommand\animageg{\adjustbox{valign=m,vspace=1pt}{\includegraphics[width=0.28\linewidth]{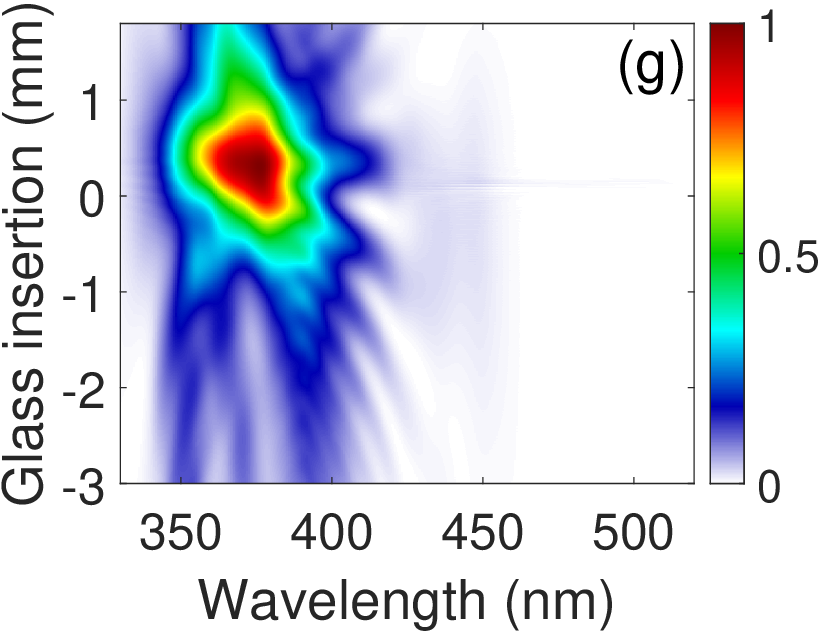}}}
\newcommand\animageh{\adjustbox{valign=m,vspace=1pt}{\includegraphics[width=0.28\linewidth]{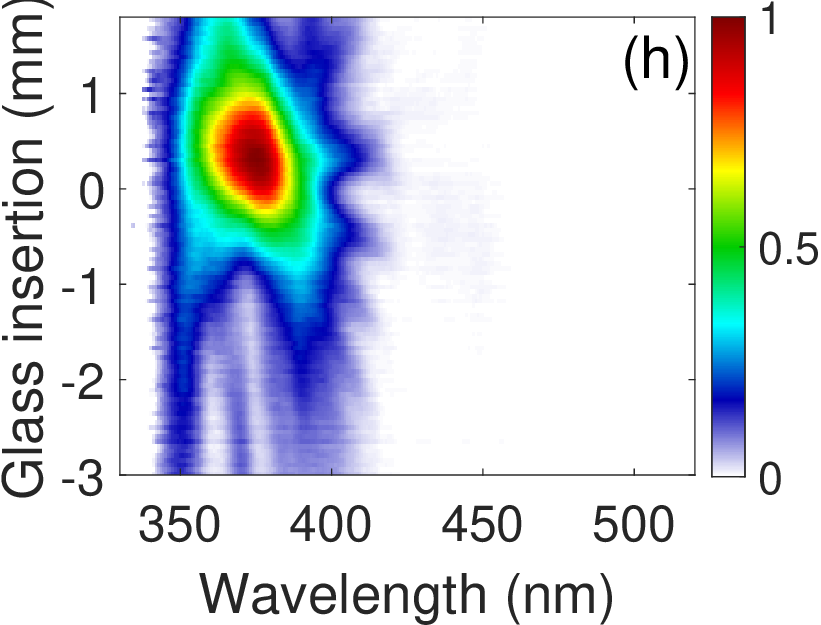}}}
\newcommand\animagei{\adjustbox{valign=m,vspace=1pt}{\includegraphics[width=0.28\linewidth]{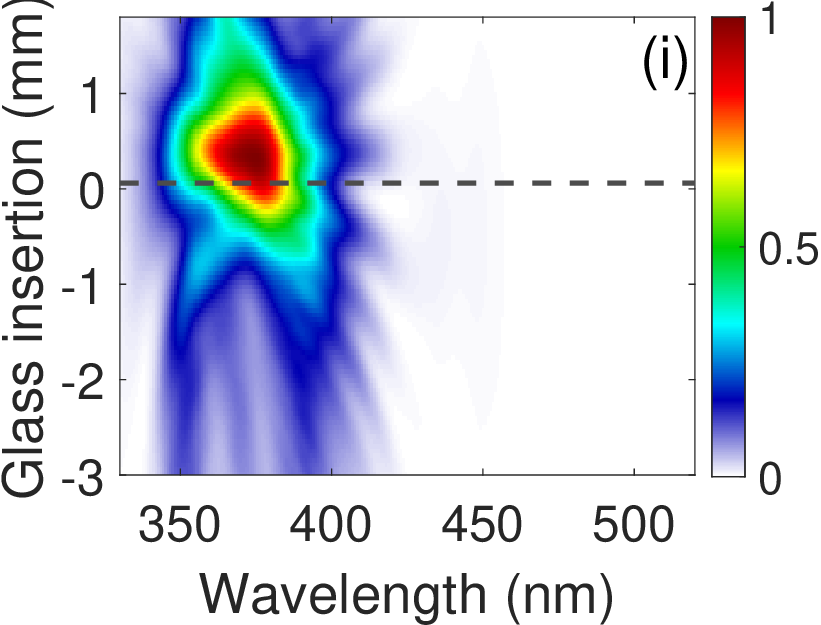}}}
\newcommand\animagej{\adjustbox{valign=m,vspace=1pt}{\includegraphics[width=0.28\linewidth]{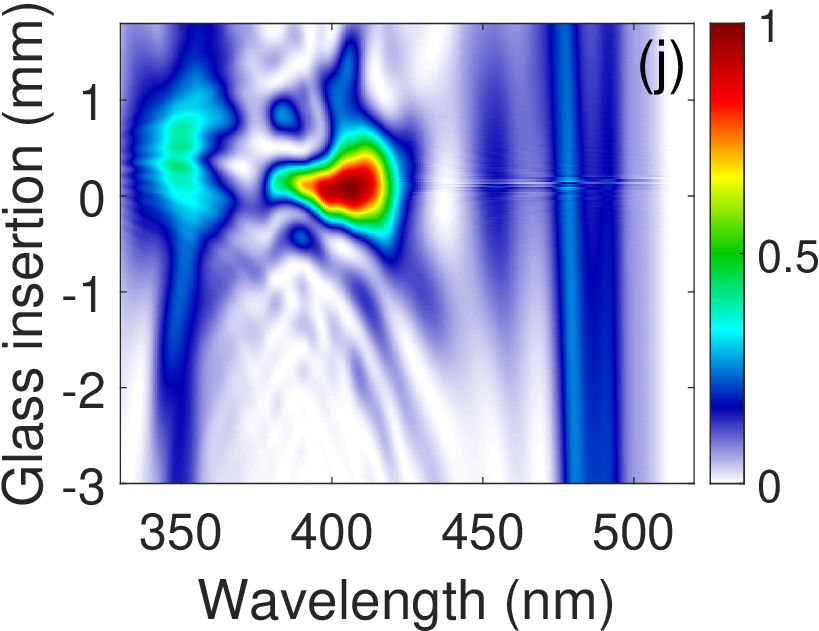}}}
\newcommand\animagek{\adjustbox{valign=m,vspace=1pt}{\includegraphics[width=0.28\linewidth]{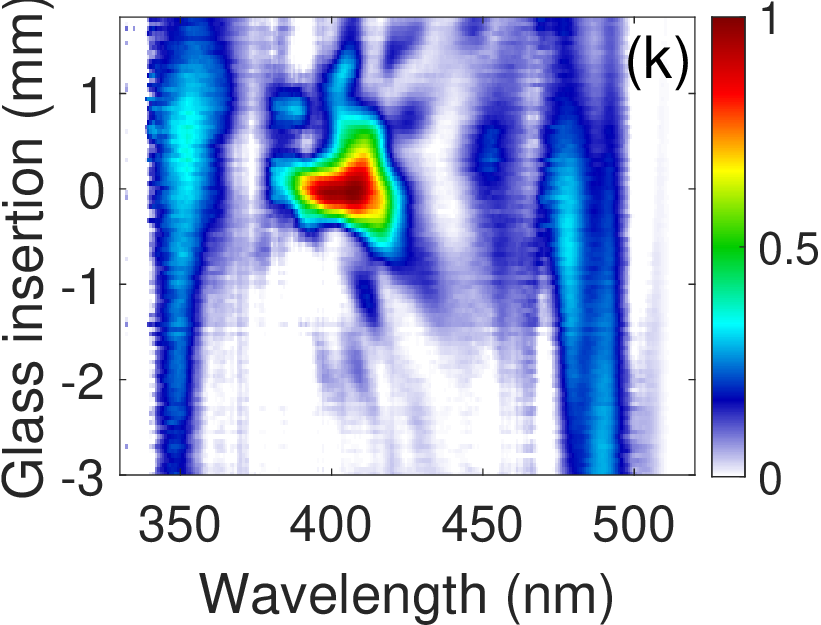}}}
\newcommand\animagel{\adjustbox{valign=m,vspace=1pt}{\includegraphics[width=0.28\linewidth]{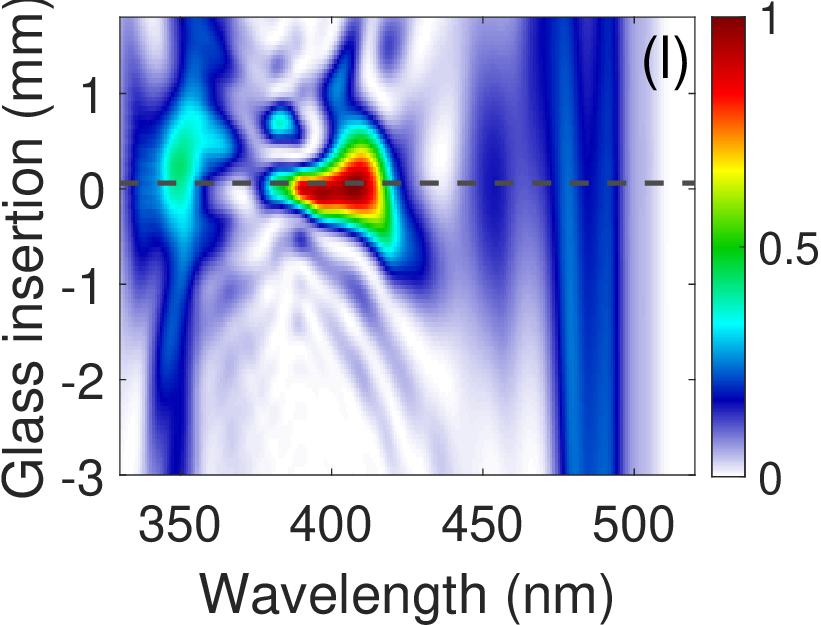}}}

\newcommand\imagea{\adjustbox{valign=m,vspace=1pt}{\includegraphics[width=0.4\linewidth]{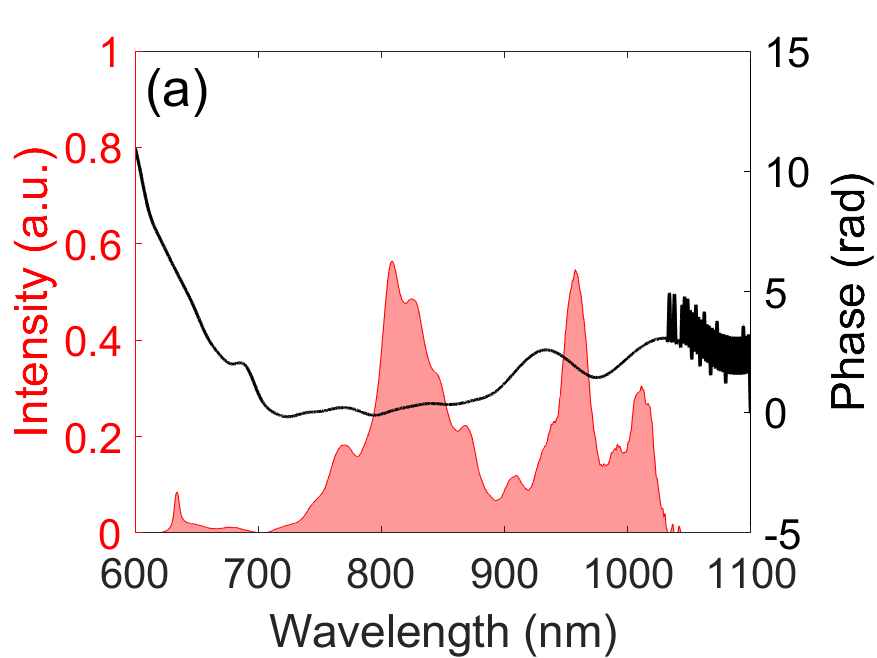}}}
\newcommand\imageb{\adjustbox{valign=m,vspace=1pt}{\includegraphics[width=0.4\linewidth]{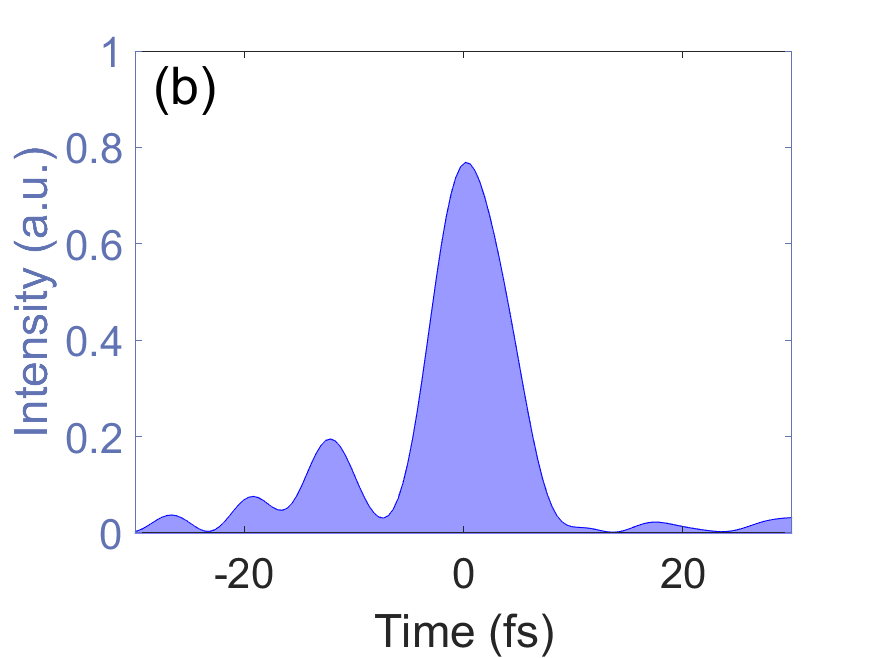}}}
\newcommand\imagec{\adjustbox{valign=m,vspace=1pt}{\includegraphics[width=0.4\linewidth]{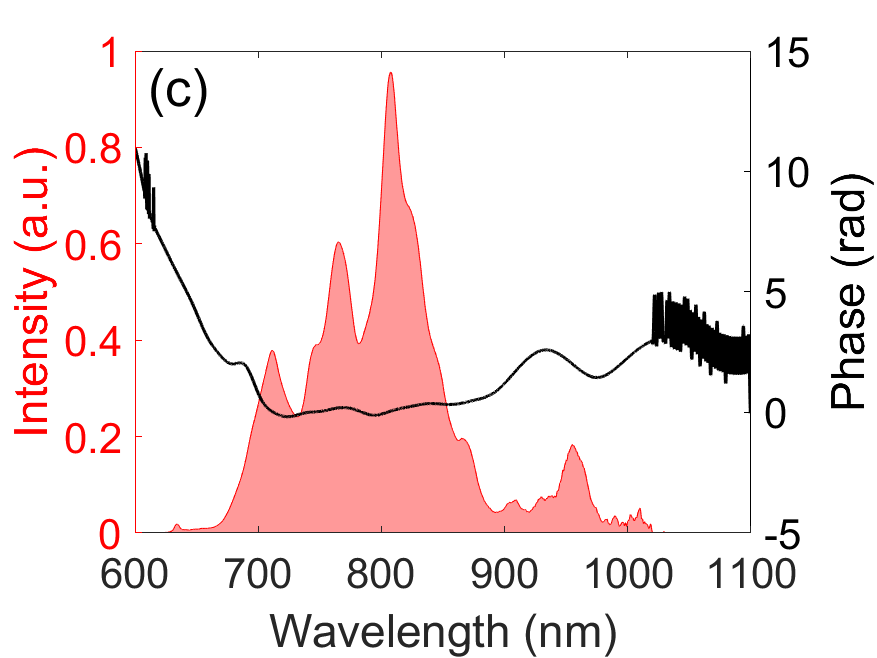}}}
\newcommand\imaged{\adjustbox{valign=m,vspace=1pt}{\includegraphics[width=0.4\linewidth]{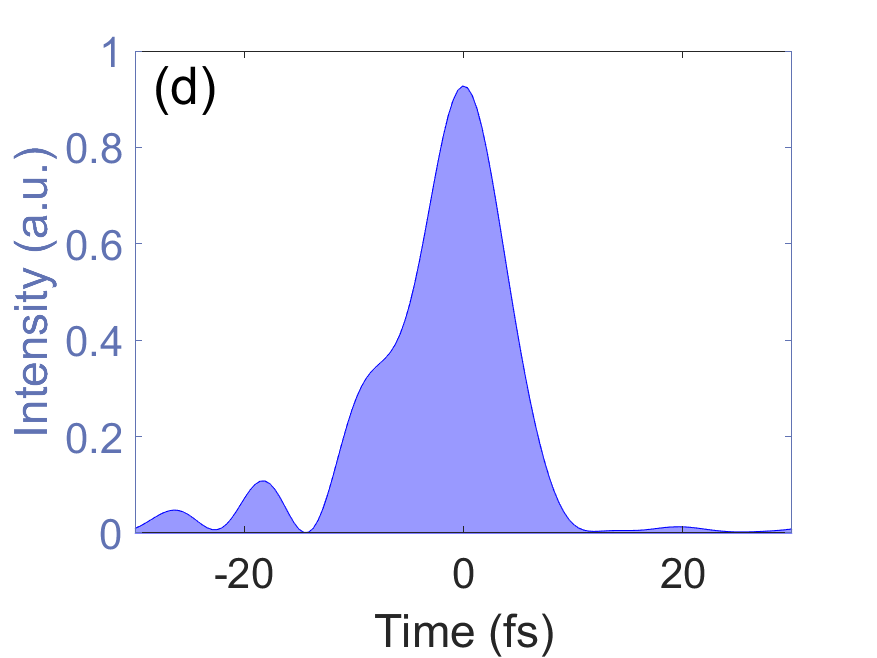}}}
\newcommand\imagee{\adjustbox{valign=m,vspace=1pt}{\includegraphics[width=0.4\linewidth]{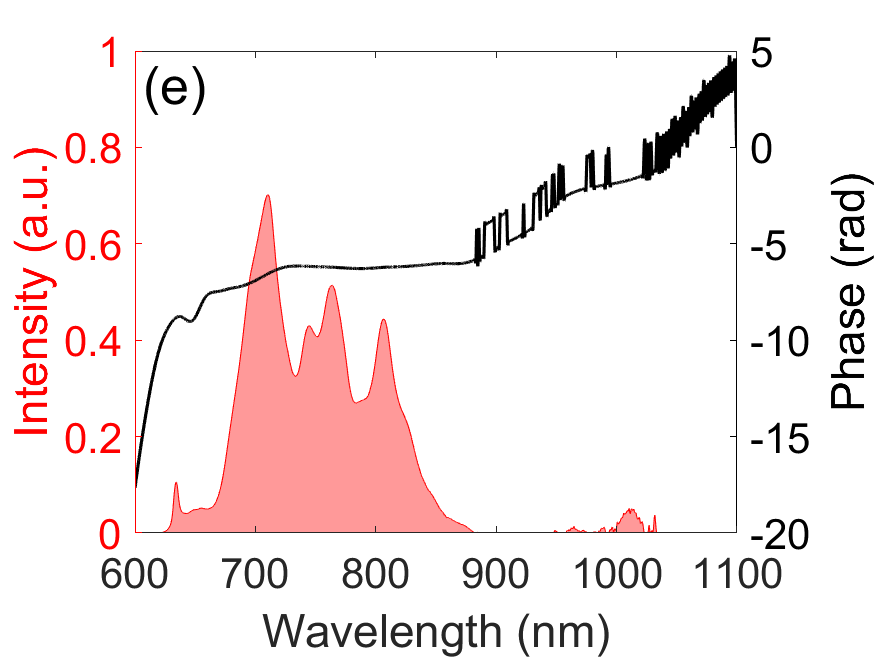}}}
\newcommand\imagef{\adjustbox{valign=m,vspace=1pt}{\includegraphics[width=0.4\linewidth]{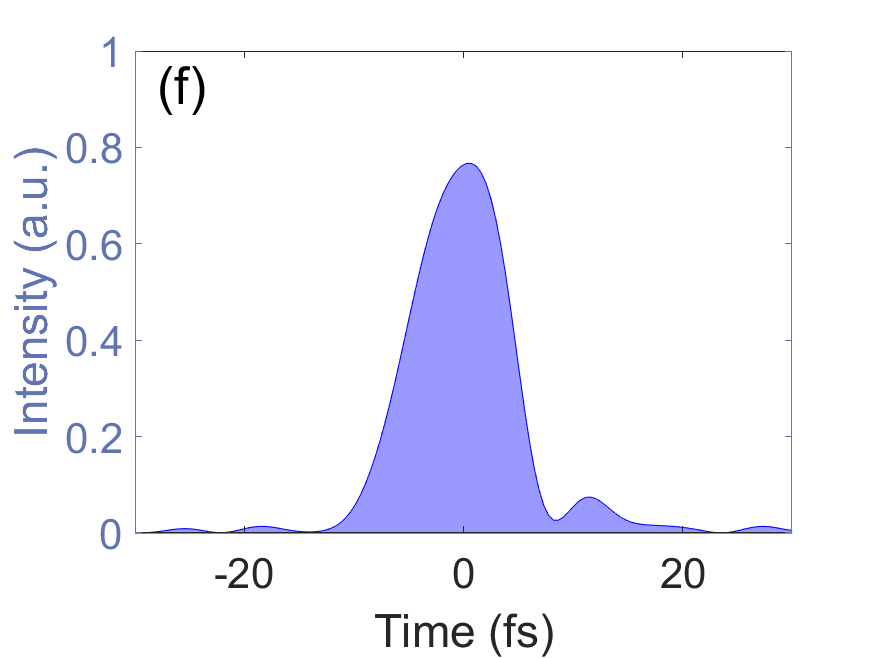}}}
\newcommand\imageg{\adjustbox{valign=m,vspace=1pt}{\includegraphics[width=0.4\linewidth]{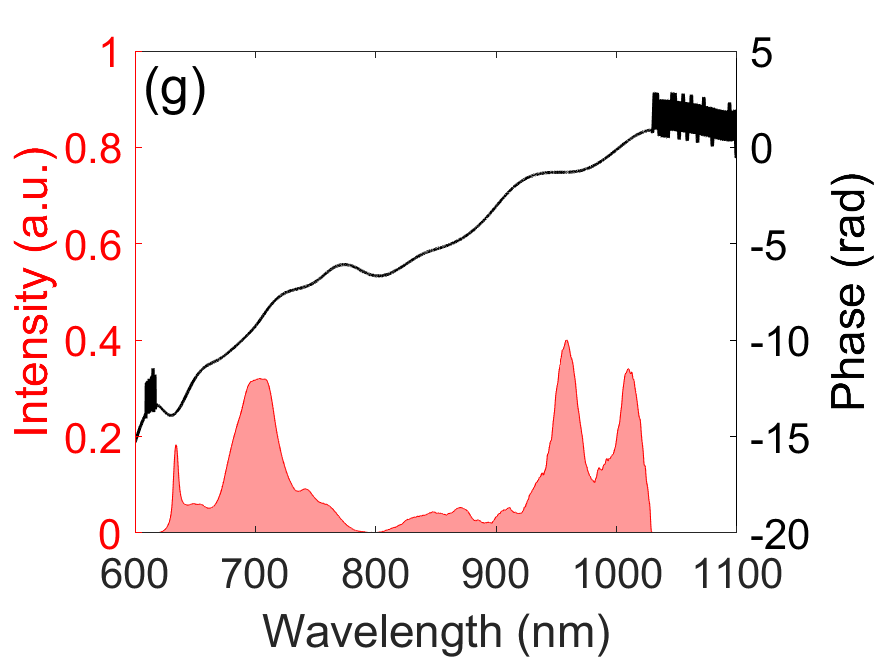}}}
\newcommand\imageh{\adjustbox{valign=m,vspace=1pt}{\includegraphics[width=0.4\linewidth]{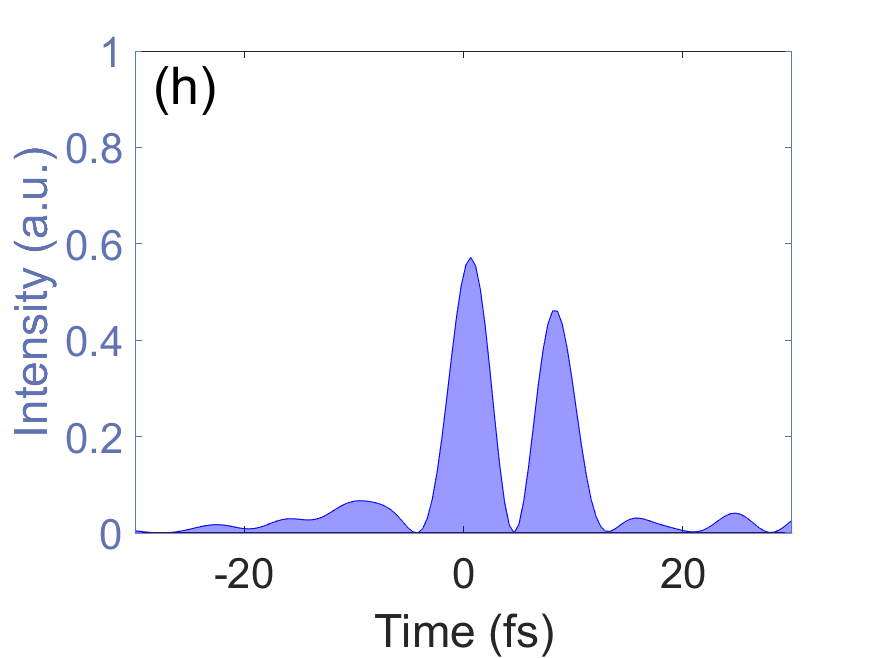}}}

\begin{document}

\title{Measurement of Ultrashort Laser Pulses With a Time-Dependent
Polarization State Using the D-Scan Technique}

\author{Daniel Díaz Rivas,\authormark{1,2,*} Ann-Kathrin Raab,\authormark{1,2} Chen Guo,\authormark{1} Anne-Lise Viotti,\authormark{1} Ivan Sytcevich, \authormark{1} Anne L'Huillier,\authormark{1} and Cord Arnold\authormark{1}}

\address{\authormark{1}Department of Physics, Lund University, P.O. Box 118, SE-22100, Lund, Sweden\\
\authormark{2}The authors contributed equally to this work.}

\email{\authormark{*}daniel.diaz\_rivas@fysik.lth.se} 


\begin{abstract*} 
The dispersion scan (d-scan) technique is extended to measurement of the time-dependent polarization state of ultrashort laser pulses. In the simplest implementation for linearly polarized ultrashort pulses, the d-scan technique records the second harmonic generation (SHG) spectrum as a function of a known spectral phase manipulation. By applying this method to two orthogonally polarized projections of an arbitrary polarized electric field and by measuring the spectrum at an intermediate angle, we can reconstruct the evolution over time of the polarization state. We demonstrate the method by measuring a polarization gate generated from $\SI{6}{fs}$ pulses with a combination of waveplates. The measurements are compared to simulations, showing an excellent agreement.

\end{abstract*}


\section{Introduction}
\label{sec:introduction}

The ability to manipulate and measure the temporal evolution of polarization states plays an important role in the control and interpretation of light-matter interactions. Femtosecond laser pulses with time-varying polarization, often referred to as polarization gates, have been utilized in quantum control for various purposes, including controlling the ionization of atoms and molecules \cite{kakehata2000, Suzuki2004, Brixner2004}, studying molecular chirality \cite{Cireasa2015, Ayuso2019}, and investigating the response of plasmonic and magnetic materials \cite{Spektor2017, Kimel2005}. Several methods have been established for generating time-varying polarization states on the femtosecond timescale. Among these are the use of Fourier-domain pulse shapers, which are based on either liquid crystals \cite{Brixner2001, weber2007} or metasurfaces \cite{chen2023}, as well as setups employing birefringent elements \cite{Altucci1998, Tcherbakoff2003, kupka2009}. Utilizing birefringent plates to modulate the degree of ellipticity as a function of time was first theoretically explored by Duncan et al. in 1990 \cite{Duncan1990}. This approach rapidly found applications, particularly in high-order harmonic generation (HHG), where the ellipticity of the fundamental driving field affects the emission efficiency of high-order harmonics significantly \cite{Budil1993}. Initial investigations employed long femtosecond pulses ($\sim \SI{150}{fs}$) \cite{Altucci1998}. Subsequent studies have extended its applicability to shorter pulses ($\sim \SI{35}{fs}$) \cite{Tcherbakoff2003}, and eventually to the few-cycle pulse regime ($\sim \SI{5}{fs}$) \cite{Sansone2006}. In the case of few-cycle pulses, the technique has enabled the generation of isolated attosecond pulses via HHG \cite{Sansone2006}, by confining the linear polarization state to a short time interval ($\sim$ a half cycle) at the center of the driving laser pulse.
\par
Characterizing the temporal evolution of these polarization states is equally important, and several techniques have been developed for this purpose. One of the pioneering methods to accomplish this is time-resolved ellipsometry (TRE) \cite{Paul1996}. This approach is challenging as one has to measure the cross-correlation with a reference pulse through a nonlinear process, for different projections of the electric field. A simpler approach is followed in the dual-channel spectral interferometry method \cite{Walecki1997}, known as POLLIWOG, where two orthogonal components of the pulse to be measured are characterized relative to a reference pulse using spectral interferometry \cite{Lepetit1995}. More recently, a method has been proposed for measuring time-dependent polarization states using in-line single-channel spectral interferometry \cite{Alonso2019,Alonso2020}. This approach employs a birefringent element to create two delayed replicas of the same pulse with orthogonal polarizations. The spectral interference analysis of the measured spectrum at a cross-angle can provide the relative phase difference between the two orthogonal projections. However, this only works well if the two projections have similar spectra. While spectral interferometric methods need a known reference pulse, implying that more than one measurement is needed to reconstruct the polarization state, a simple non-iterative algorithm can be effectively implemented to determine the relative phase between orthogonal components of the electric field. Very recently, it has been shown that it is possible to reconstruct the polarization state from a single measurement with the amplitude-swing technique \cite{barbero2023}, which was initially developed to measure the spectral phase of linearly polarized pulses \cite{Aswing}. 
\par
 An alternative approach is using a standard method for characterizing scalar pulses, i.e., pulses with constant linear polarization. Due to the insensitivity of most of these methods to the zero- and first-order terms of the spectral phase, it is not enough to measure two orthogonally polarized components of the pulse to reconstruct its polarization. The tomographic ultrafast retrieval of transverse light E-fields (TURTLE) method addresses this issue by reconstructing the polarization state from three distinct projections \cite{Schlup2008, Xu2009}. Recently, this method has been applied to reconstruct the pulse from a single measurement. Instead of measuring three projections, the angle of the polarizer is rotated as the pulse is characterized \cite{Haham2021}.
\par
\begin{figure}[H]
\centering
\includegraphics[width=0.6\textwidth]{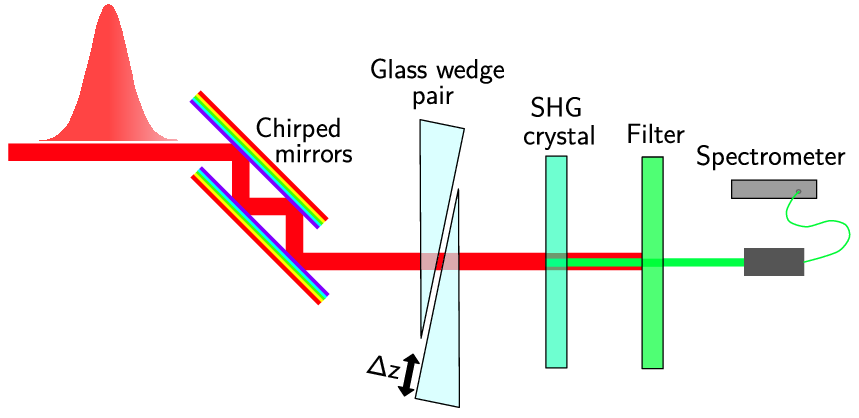}
\caption{Simplified schematic of a d-scan setup.}
\label{fig:conventional_dscan}
\end{figure}

The characterization of ultrashort laser pulses with linear polarization has advanced significantly over the past few decades. Widely used techniques include autocorrelation \cite{Sala1980}, frequency-resolved optical-gating (FROG) \cite{Trebino1993}, and spectral interferometry for direct electric-field reconstruction (SPIDER) \cite{Iaconis1999}. The dispersion scan (d-scan) technique has emerged as a promising alternative in recent years \cite{Sytcevich2021}, often being the preferred method for laser pulse characterization in the few-cycle regime. A schematic of a typical d-scan implementation is illustrated in Fig.~\ref{fig:conventional_dscan}.
\par
A d-scan measurement involves transmitting laser pulses through a pair of glass wedges to introduce positive dispersion, predominantly in the form of group delay dispersion (GDD). Chirped mirrors can be employed to ensure that the pulses reach the wedges with a negative chirp. At a specific glass insertion, the wedge pair compensates for the negative GDD, yielding approximately transform-limited pulses. The pulses subsequently pass through a nonlinear material, producing a signal such as second harmonic generation (SHG), which is measured using a spectrometer. By scanning the wedges around the optimal compression point while recording the nonlinear signal spectrum, a 2D spectrogram called the d-scan trace is obtained, which can be easily simulated and provides intuitive information about the pulse. As with FROG, a phase retrieval algorithm is required to extract the phase from the d-scan trace by iteratively minimizing the error between the simulated and measured traces. The d-scan technique provides a straightforward in-line setup that eliminates the need for a reference pulse or interferometric precision. Additionally, its compatibility with dispersion elements commonly found in ultrafast laser compressor systems allows for simultaneous compression and pulse measurement.
\par
In this work, we extend d-scan to the measurement of time-dependent polarization states. The proposed method consists in measuring the amplitude and phase of two orthogonal projections of the electric field. The spectral interference analysis of the measured spectrum at a cross-angle provides the relative phase difference between the two orthogonal projections. We apply this method to the characterization of a polarization gate in the few-cycle regime.
\par
Following this introduction, the methods section presents a description of the polarization gate generation and its characterization using the d-scan technique. The results are then presented and discussed.

\section{Methods}
\label{sec:methods}

In this section, the experimental setup and methodologies used for creating and characterizing a time-dependent polarization state from a few-cycle linearly polarized laser pulse are detailed.

\subsection{From Constant to Time-Dependent Polarization}
\label{sec:polarization_gate}

\begin{figure}[h]
\centering
\begin{subfigure}{0.30\textwidth}
    \centering
    \includegraphics[width=\linewidth]{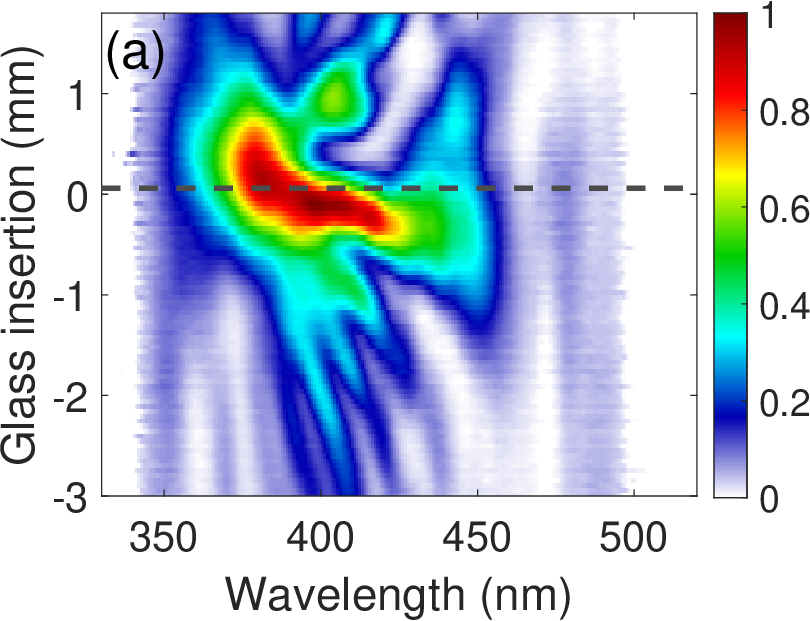}
\end{subfigure}
\begin{subfigure}{0.32\textwidth}
    \centering
    \includegraphics[width=\linewidth]{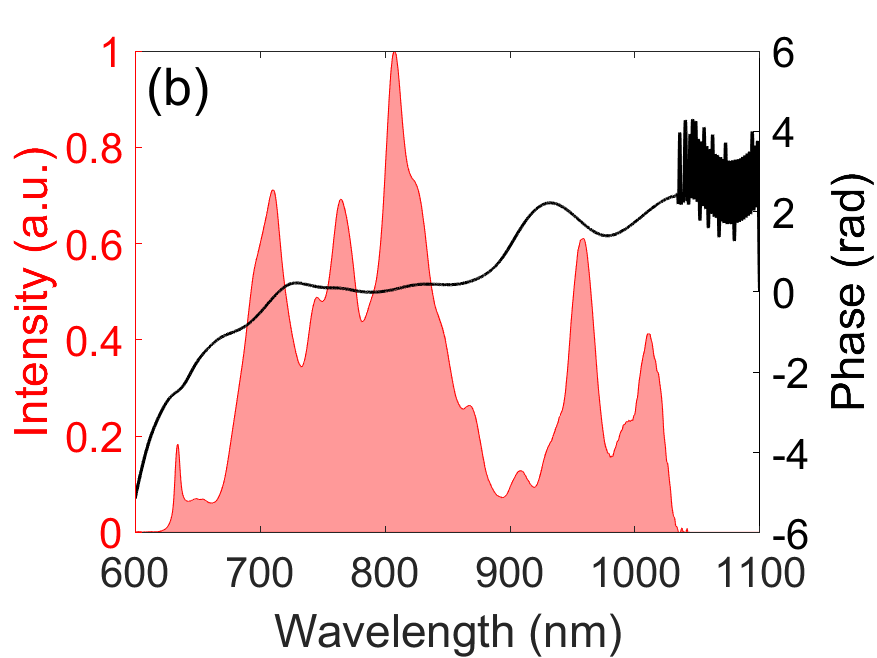}
\end{subfigure}
\begin{subfigure}{0.32\textwidth}
    \centering
    \includegraphics[width=\linewidth]{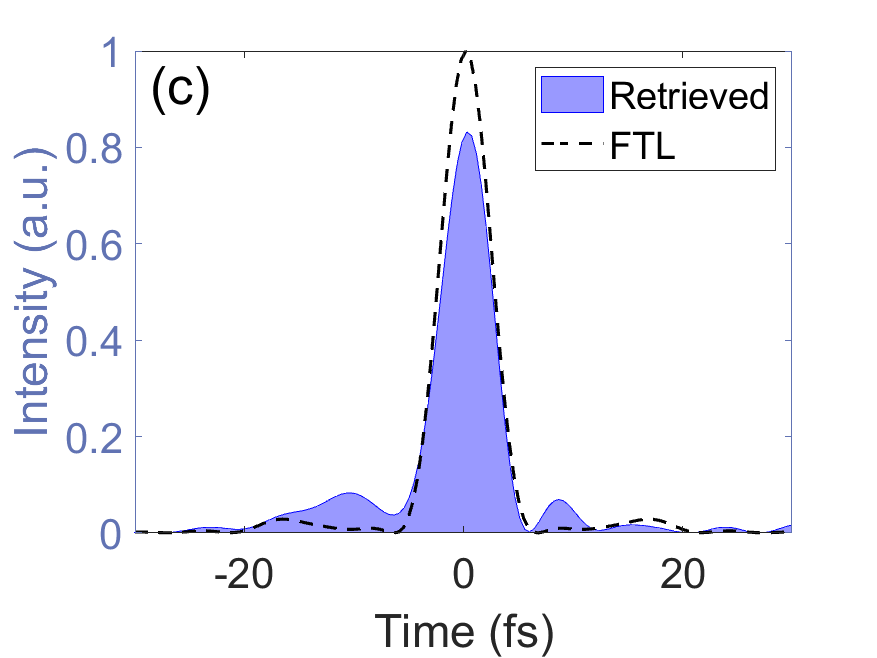}
\end{subfigure}
\caption{Characterization of the few-cycle laser pulses utilized in the experiments: (a) measured d-scan trace; (b) experimental spectrum and retrieved spectral phase at optimal compression; (c) temporal intensity of the retrieved pulse at optimal compression. Optimal compression is achieved at the position denoted by the dashed line in the measured trace.}
\label{fig:input_pulse_characterization}
\end{figure}

In the experiments conducted for this study, few-cycle laser pulses are generated by an Optical Parametric Chirped Pulse Amplification (OPCPA) source \cite{Harth2017}. These pulses, centered around a wavelength of approximately $\SI{850}{nm}$, exhibit an average power of $\SI{3}{W}$ with a repetition rate of $\SI{200}{kHz}$. Characterization of these pulses is accomplished using the d-scan technique, as shown in Fig.~\ref{fig:input_pulse_characterization}.
\par
In Fig.~\ref{fig:input_pulse_characterization}(a), the "zero" insertion denotes the insertion point at which the pulse is the shortest. This corresponds to the spectral phase in Fig.~\ref{fig:input_pulse_characterization}(b) and pulse profile in Fig.~\ref{fig:input_pulse_characterization}(c). At the point of optimal compression, the pulse reconstruction yields a full-width at half-maximum (FWHM) of \SI{5.5}{fs}. A comparison with the Fourier-limited pulse, derived from the measured spectrum and depicted by the dashed line in Fig.~\ref{fig:input_pulse_characterization}(c), shows that approximately $80\%$ of the Fourier-limited pulse peak intensity is achieved at the position of optimal compression.
\par
The pulses emitted from the OPCPA are initially linearly polarized. A time-dependent polarization state, which is the focus of this study, is then created by utilizing a combination of waveplates. The process begins with a quartz birefringent plate that acts as a third-order quarter-wave plate (1.75$\lambda$) for a wavelength of $\SI{800}{nm}$. This causes the pulse to split into two orthogonally polarized components, delayed by approximately $\SI{4.7}{fs}$, thereby inducing a time-dependent polarization state. The polarization changes from linear to elliptical, becomes circular when the components are of equal amplitude, and then reverses back to linear as the amplitudes vary. Although the first plate alone generates a time-dependent polarization state, applications in attosecond science benefit from a confined linear polarization state at the pulse center. For this, the pulses are then transmitted through a second quartz plate acting as a zero-order quarter-wave plate (0.25$\lambda$) at $\SI{800}{nm}$. This second plate is set at a $45^\circ$ angle relative to the first one, which enables the conversion of circular polarization to linear polarization, and vice versa.
\par
\begin{figure}[H]
\centering
\includegraphics[width=0.8\textwidth]{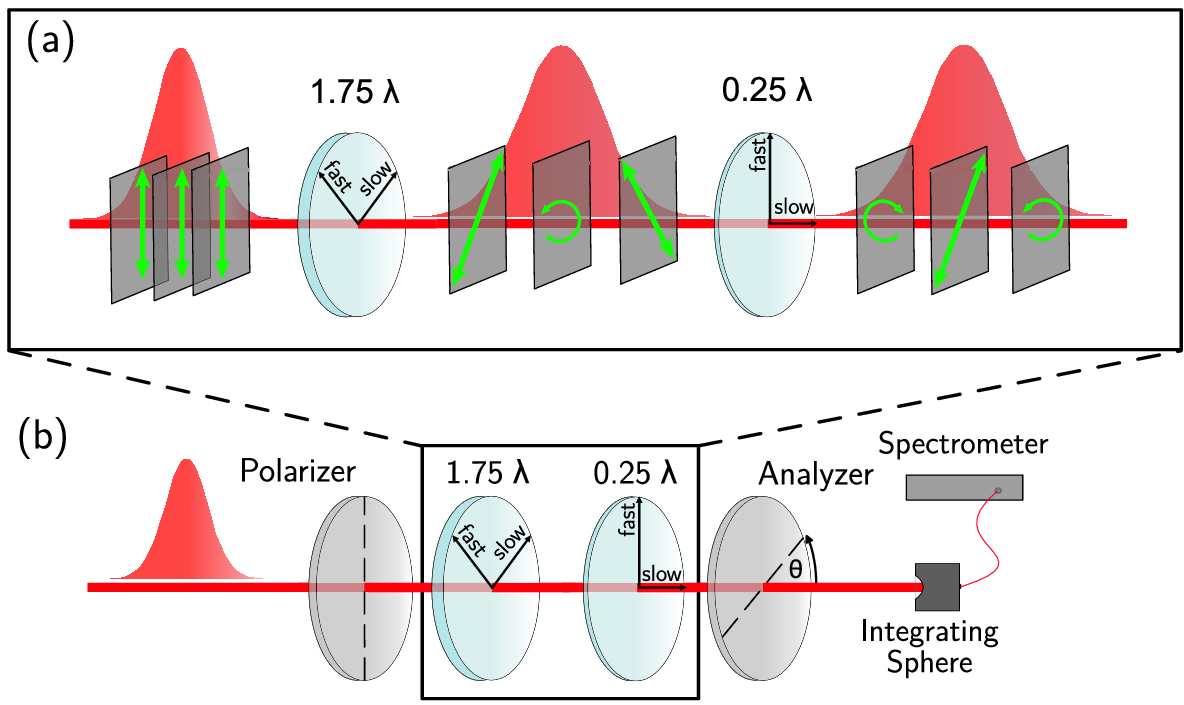}
\caption{Schematic diagram illustrating: (a) the generation process of the polarization gate; (b) the setup for spectral characterization of the polarization gate, which includes placing the polarization gate generation between a pair of polarizers, and measuring the transmitted spectrum as the second polarizer ("analyzer") is rotated.}
\label{fig:polarization_gate_setup}
\end{figure}

This scheme is illustrated in Fig.~\ref{fig:polarization_gate_setup}(a). The polarization evolves from circular polarization, to linear in the middle, and returns to circular polarization with opposite handedness.
\par

\subsection{Characterization of the Polarization Gate}

\subsubsection{Spectral Characterization}
\label{sec:spectral_characterization}

To characterize the polarization gate, a simple setup is constructed to examine the polarization state as a function of wavelength, which is depicted in Fig.~\ref{fig:polarization_gate_setup}(b). Initially, the few-cycle laser pulses pass through a vertically oriented linear polarizer to ensure a clean vertical polarization. Subsequently, the two quarter-wave plates discussed in the previous section are introduced to generate the polarization gate. A second polarizer, referred to as the "analyzer," can be rotated to project the electric field at any angle. The transmitted light is collected by an integrating sphere, which transmits the signal to a fiber-coupled spectrometer (AvaSpec-3648, from AvantesInc) capable of operating within the spectral range of $\SI{200}{nm}$ to $\SI{1100}{nm}$.
\par
During the characterization process, measurements are taken as the analyzer is rotated, in order to measure the spectrum corresponding to different projections of the electric field. As a measurement convention, an analyzer angle of $0^\circ$ corresponds to the initial position, where its polarization axis is aligned with the horizontal direction. The spectrum is recorded every $10^\circ$ as the analyzer is rotated from $0^\circ$ to $180^\circ$, which corresponds to one complete revolution of its polarization axis.

\subsubsection{Temporal Characterization}
\label{sec:polarization_dscan}

\begin{figure}[H]
\centering
\includegraphics[width=0.8\textwidth]{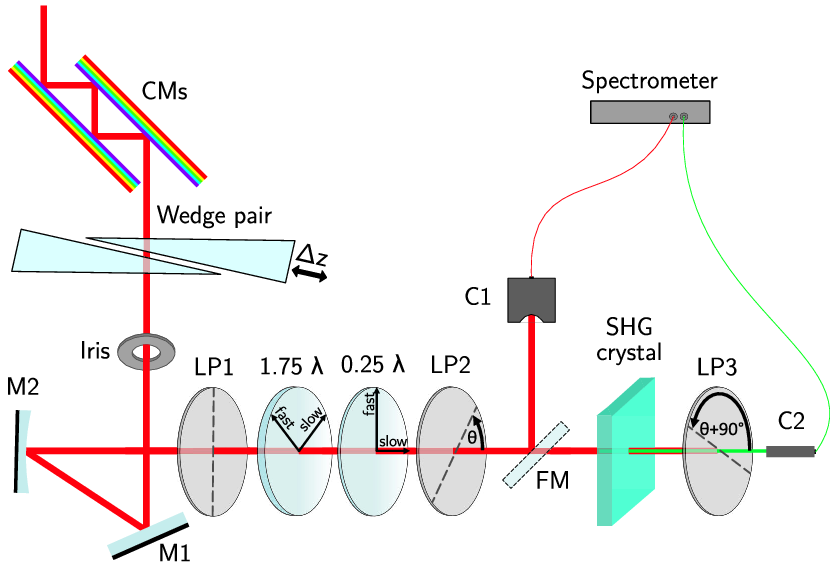}
\caption{Schematic diagram of the polarization d-scan setup. In the diagram, there are some abbreviations to represent various optical elements. 'M' denotes Metallic Mirror, 'FM' for Flip Mirror, 'LP' for Linear Polarizer, and 'C' for Light Collector.}
\label{fig:polarization_dscan_setup}
\end{figure}

We call the measurement of a time-dependent polarization state using the d-scan technique polarization d-scan. A drawing of the setup is presented in \autoref{fig:polarization_dscan_setup}. This setup has two main functions: to include the optical elements necessary for the generation of the polarization gate, and to measure linearly polarized projections of the electric field using the d-scan technique.
\par
The dispersion of the pulses is controlled using broadband chirped mirrors and a BK7 glass wedge pair. Subsequently, the spectral characterization setup is integrated, which both produces the desired time-dependent polarization state and transmits different projections of the polarization gate. A BBO crystal is placed in the focal plane of mirror M2 to generate a second harmonic signal, while an additional linear polarizer (LP3) filters out the fundamental radiation and any residual second harmonic present before the nonlinear crystal. For each projection, both the fundamental spectrum (via a flip mirror) and SHG spectrum are measured. This setup enables the d-scan measurement of distinct linearly polarized projections of the polarization gate pulse. Measurements are ultimately processed with a homemade d-scan software that retrieves the spectral phase of the pulse and reconstructs the pulse profile in the time domain.

\subsection{Reconstruction of the Polarization State}
\label{sec:reconstruction_polarization_state}

From the polarization d-scan, one can obtain the amplitude and phase of two orthogonal projections of the electric field, such as $E_{0^\circ}$ and $E_{90^\circ}$:
\par
\begin{align}
    E_{0^\circ}(\nu)& = \left|E_{0^\circ}(\nu)\right| \exp{\left(i\phi_{0^\circ}(\nu)\right)} \label{eq:e0}\\
    E_{{90}^\circ}(\nu)& = \left|E_{{90}^\circ}(\nu)\right| \exp{\left(i\phi_{{90}^\circ}(\nu)\right)}  \label{eq:e90}
\end{align}
\par
In general, the polarization state can be calculated as the sum of two orthogonal components of the electric field. However, since the d-scan technique is insensitive to the first and zero order terms of the spectral phase, the retrieval of two orthogonal pulse components using the d-scan technique is insufficient to reconstruct the time-dependent polarization state. To accommodate this, an expression can be written for the electric field $\mathbf{E}(\nu)$ where a phase factor is included to account for the first and zero order terms of the phase:
\par
\begin{equation}
    \mathbf{E}(\nu)= E_{0^\circ}(\nu) \hat{\mathbf{e}}_\mathbf{{0^\circ}} + E_{{90}^\circ}(\nu) \exp{\left[-i(2\pi\nu \tau+\varphi)\right]} \hat{\mathbf{e}}_\mathbf{{90^\circ}}
\label{eq:electric_field}
\end{equation}
\par
Each component is represented by the product of the electric field complex amplitude, obtained from the d-scan, and the unit vector in the respective direction $\left(\hat{\mathbf{e}}_\mathbf{{0^\circ}}, \hat{\mathbf{e}}_\mathbf{{90^\circ}}\right)$. The phase factor multiplies the second component and consists of a first order ($\tau$) and zero order ($\varphi$) phase terms. From this expression, the electric field at an arbitrary angle $\alpha$ can be computed using:
\par
\begin{equation}
    E_{\alpha}(\nu)= \cos{\left(\alpha\right)} E_{0^\circ}(\nu) + \sin{\left(\alpha\right)} E_{{90}^\circ}(\nu) \exp{\left[-i(2\pi\nu \tau+\varphi)\right]}
\label{eq:formula_efield_alpha}
\end{equation}
\par
The transmitted spectrum at an angle $\alpha$, $I_\alpha(\nu)$, depends on $\varphi$ and $\tau$, as well as on the retrieved spectral phase for each projection, $\phi_{0^\circ}$ and $\phi_{90^\circ}$.
\begin{multline}
    I_{\alpha}(\nu) = \cos{\left(\alpha\right)}^2 |E_{0^\circ}(\nu)|^2 + \sin{\left(\alpha\right)}^2|E_{90^\circ}(\nu)|^2\\ + 2 \cos{\left(\alpha\right)} \sin{\left(\alpha\right)} |E_{0^\circ}(\nu)| |E_{90^\circ}(\nu)| \cos{\left(\phi_{0^\circ}(\nu) - \phi_{90^\circ}(\nu) + 2\pi\nu \tau + \varphi \right)}
\label{eq:formula_intensity_alpha}
\end{multline}
\par
In practice, the roles are reversed, as $\tau$ and $\varphi$ are the unknowns while the transmitted spectrum at an angle $\alpha$ can be simply obtained from a measurement of the fundamental spectrum for a projection at this angle. The method presented in this work uses an optimization algorithm to identify the values of $\tau$ and $\varphi$ that minimize the difference between the retrieved spectrum at the angle $\alpha$ and the calculated one based on Eq.~\ref{eq:formula_intensity_alpha}. The algorithm defines lower and upper bounds for $\tau$ and $\varphi$ as $\SI{-10}{fs} < \tau < \SI{10}{fs}$ and $-\pi < \varphi < \pi$, respectively. Then, a grid of $100 \times 100$ equidistantly  points is constructed within these boundaries. For each point on the grid, the root mean square error (RMSE) is calculated to estimate the error:
\par
\begin{equation}
    \text{RMSE} = \sqrt{\frac{\sum{(I_{\alpha}^{\text{meas}}-I_{\alpha}^{\text{sim}})^2}}{N}}
\label{eq:RMSE}
\end{equation}
\par
Here, $I_{\alpha}^{\text{meas}}$ and $I_{\alpha}^{\text{sim}}$ are the measured and simulated spectra at the angle $\alpha$, respectively, and $N$ is the number of data points. The optimization algorithm seeks the values of $\tau$ and $\varphi$ that minimize the RMSE across the grid points.
\par
The electric field can then be calculated using Eq.~\ref{eq:electric_field}. To obtain the evolution of the polarization state over time, the computed field can be Fourier transformed. From the electric field one can also calculate the envelope of the polarization gate pulse ($I(t)$), using:
\par
\begin{equation}
    I(t) = I_{0^\circ}(t) + I_{90^\circ}(t) =  |E_{0^\circ}(t)|^2 + |E_{90^\circ}(t)|^2
\label{eq:envelope}
\end{equation}

\subsection{Modelling of the Polarization Gate}
\label{sec:simulations}

A theoretical model employing the Jones matrix formalism for broadband pulses can be utilized to simulate the polarization gate pulse and its characterization process. Initially, the pulses have a linearly polarized state denoted by the Jones vector $\mathbf{J_{i}}$. The non-zero component of this vector corresponds to the complex amplitude of the pulse in the frequency domain. The evolution of the Jones vector $\mathbf{J_{f}}$ through the combination of waveplates is calculated using matrix multiplication,
\par
\begin{equation}
    \mathbf{\tilde{J}_f} (\pmb{\nu}) = \mathbf{M_{QW0}} \times \mathbf{R}(-45^\circ) \times  \mathbf{M_{QW3}} \times \mathbf{R}(45^\circ) \times \mathbf{\tilde{J}_i}(\pmb{\nu})
\label{eq:matrix_multiplication}
\end{equation}
\par
$\mathbf{M_{QW0}}$ and $\mathbf{M_{QW3}}$ denote the Jones matrices of the zeroth-order and third-order waveplates, respectively. Furthermore, $\mathbf{R}(-45^\circ)$ and $\mathbf{R}(45^\circ)$ represent the matrices for the rotation of the frame of reference by $\SI{-45}{^\circ}$ and $\SI{45}{^\circ}$, respectively.
\par
The Jones vector in the time domain can be calculated applying the Fourier transform over the Jones vector in the frequency domain:
\par
\begin{align}
    \mathbf{J_f} (\mathbf{t}) = \mathcal{F} \left\{(\mathbf{\tilde{J}_f} (\pmb{\nu})\right\} &= \begin{bmatrix}
       A_x(\mathbf{t}) \, e^{i\varphi_x (\mathbf{t})}\\
       A_y(\mathbf{t}) \, e^{i\varphi_y (\mathbf{t})}\\
    \end{bmatrix}
\label{eq:time_jones_vector}
\end{align}
\par
The amplitude ratio $\left(\mathcal{R} = A_x/A_y\right)$ and the relative phase difference $(\Delta\varphi = \varphi_x - \varphi_y)$ between the two orthogonal components of the electric field can be determined as a function of time. Additional useful parameters for describing polarization are the angle of ellipticity $(\chi)$ and the degree of ellipticity $(\varepsilon)$ defined as \cite{photonics}:
\par
\begin{align}
    \sin{2\chi}&=\frac{2\mathcal{R}}{1+\mathcal{R}^2} \sin{\Delta\varphi} \label{eq:polarization_ellipse_eq1}\\
    \varepsilon &= \left|\tan{\chi}\right|  \label{eq:polarization_ellipse_eq2}
\end{align}
\par
The effect of a polarizer on the system is evaluated by the projection of the Jones vector along a specific direction. For the modeling of polarization d-scan measurements, a matrix is constructed from the projection of the electric field. This matrix encodes both the glass insertion and the spectrum. Each column represents a different frequency point in the spectrum, while every row corresponds to a different glass insertion point. By applying the Sellmeier equations, the dispersion associated with the BK7 wedges is computed numerically. The second harmonic spectrum is then calculated for each row of the matrix, effectively simulating the impact of different glass insertion points. This results in a matrix that forms the d-scan trace of a specific projection of the electric field, defined by the orientation of the polarizer.

\section{Results}
\label{sec:results}

This section presents and discusses the characterization of the polarization gate and the reconstruction of the time-dependent polarization state.

\subsection{Spectral Characterization of the Polarization Gate}

\begin{figure}[H]
\centering
\includegraphics[width=0.7\textwidth]{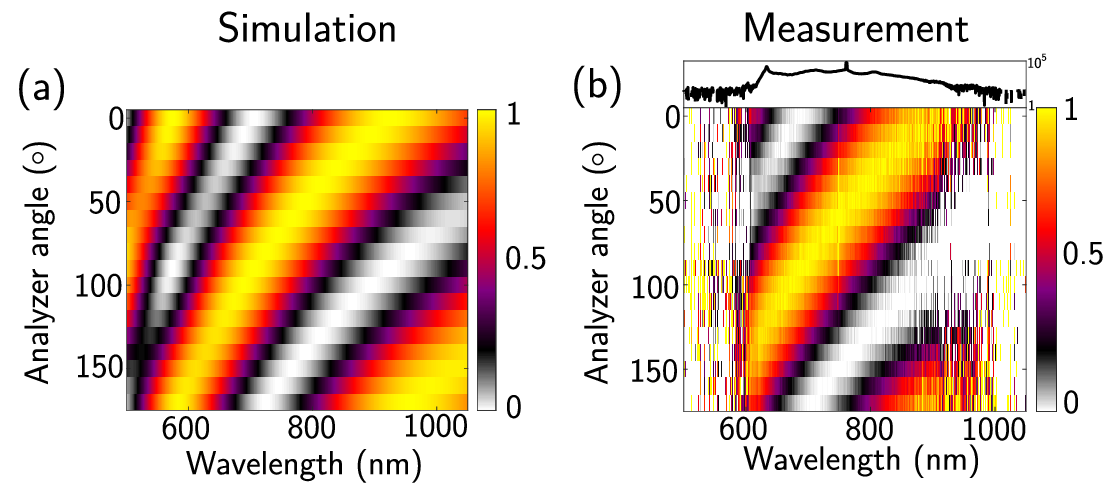}
\caption{2D plot (transmitted spectrum for different analyzer angles) illustrating the spectral characterization of the polarization gate: (a) simulation; (b) measurement. The laser spectrum, represented on a logarithmic scale, is overlaid on top of the measurement.}
\label{fig:waveplate_characterization_gate}
\end{figure}

Section~\ref{sec:spectral_characterization} outlines the method used to investigate the polarization state in the frequency domain. The spectra obtained for each wavelength are normalized by the input laser's intensity at that wavelength. Figure~\ref{fig:waveplate_characterization_gate} compares the simulated (a) and measured (b) polarization gate over the spectrum of the pulse. Each pixel in these images has a value corresponding to the light intensity for each wavelength and analyzer angle, as represented in the color map.
\par
A strong agreement between simulated and experimental results is evident. High contrast as a function of polarizer angle is apparent across the entire spectrum, indicating that the polarization state associated with each wavelength is approximately linear. The change in the position of the intensity maximum at different angles of the analyzer can also be observed when examining different wavelengths, suggesting that the direction of polarization rotates with wavelength. It is important to note that the measured signal-to-noise ratio declines rapidly outside the wavelength range of \SI{600}{nm} and \SI{1000}{nm}. This window corresponds to the actual laser spectrum, as seen in the overlaid log-scale spectrum in Fig~\ref{fig:waveplate_characterization_gate}(b).

\subsection{Temporal Characterization of the Polarization Gate with the D-Scan}

\begin{figure}[H]
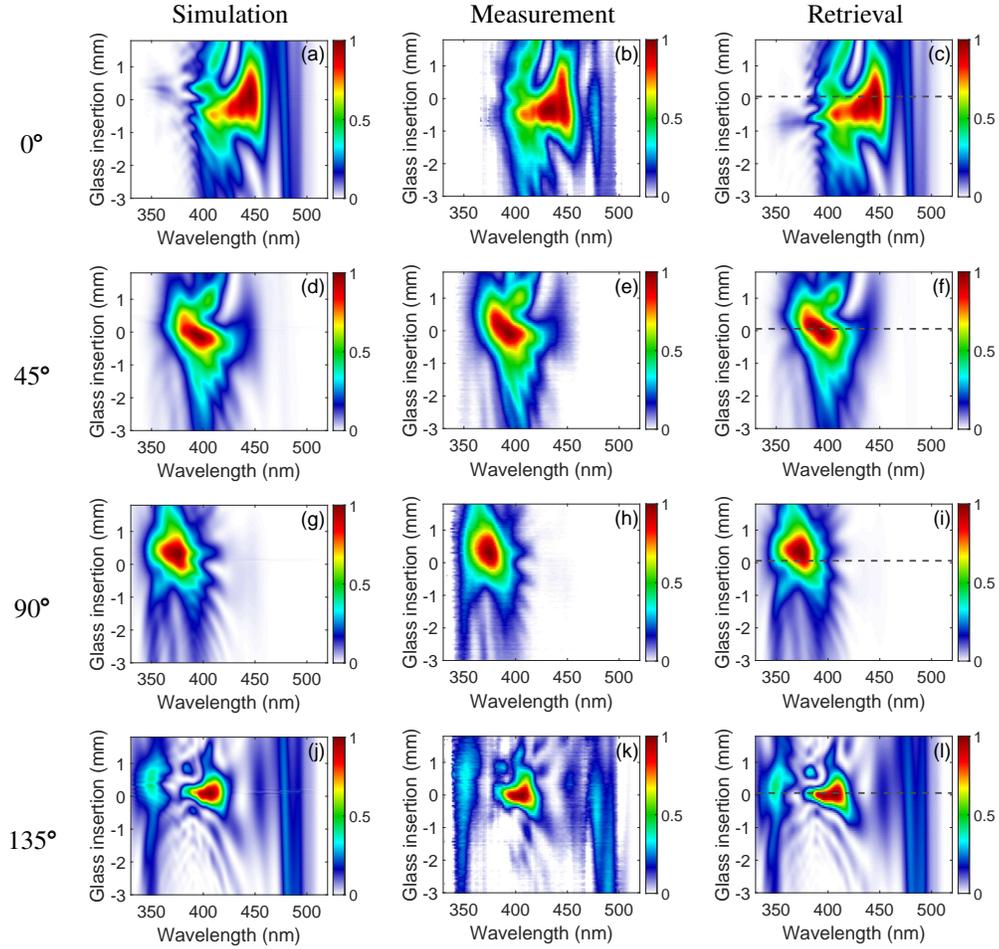

\centering
\renewcommand{\arraystretch}{1}
\begin{tabular}{cccc}
       & Simulation & Measurement & Retrieval \\
$0^{\pmb{\circ}}$  & \animagea & \animageb & \animagec \\
$45^{\pmb{\circ}}$  & \animaged & \animagee & \animagef \\
$90^{\pmb{\circ}}$  & \animageg & \animageh & \animagei \\
$135^{\pmb{\circ}}$ & \animagej & \animagek & \animagel
\end{tabular}
\renewcommand{\arraystretch}{1.0}
\caption{Simulated, measured, and retrieved d-scan traces of four linearly polarized projections of the electric field: (a)-(c) $0^\circ$; (d)-(f) $45^\circ$; (g)-(i) $90^\circ$; (j)-(l) $135^\circ$.}
\label{fig:dscan_measurements}
\end{figure}

The time-dependent polarization state is examined using the polarization d-scan method outlined in Section~\ref{sec:polarization_dscan}. Four d-scan measurements are conducted by varying the orientation of the second linear polarizer in the experimental setup (see Fig.~\ref{fig:polarization_dscan_setup}), so that each measurement corresponds to a distinct projection of the electric field. Figure~\ref{fig:dscan_measurements} displays the simulated, measured, and retrieved d-scan traces for each case.
\par
Again, a strong agreement between the simulations, measurements, and retrievals is observed, demonstrating the robustness of the simulation and the retrieval algorithm. Comparing the results between the different projections reveals significant variations in spectral content. This can be explained by examining Fig.~\ref{fig:input_pulse_characterization} and Fig.~\ref{fig:waveplate_characterization_gate}. The second harmonic spectrum is generated by doubling the frequency of the fundamental spectrum, shown in Fig.~\ref{fig:input_pulse_characterization}(b). As a result, only wavelengths within this window contribute to the generation of the second harmonic signal. Moreover, each measurement in Fig.~\ref{fig:dscan_measurements} corresponds to a different projection of the electric field. As illustrated in Fig.~\ref{fig:waveplate_characterization_gate}, the intensity distribution of the fundamental spectrum after the second polarizer depends on its orientation. Since different wavelengths of the spectrum are associated with different polarization states, various projections made with the polarizer result in different parts of the spectrum being selected.
\par
\begin{figure}[H]
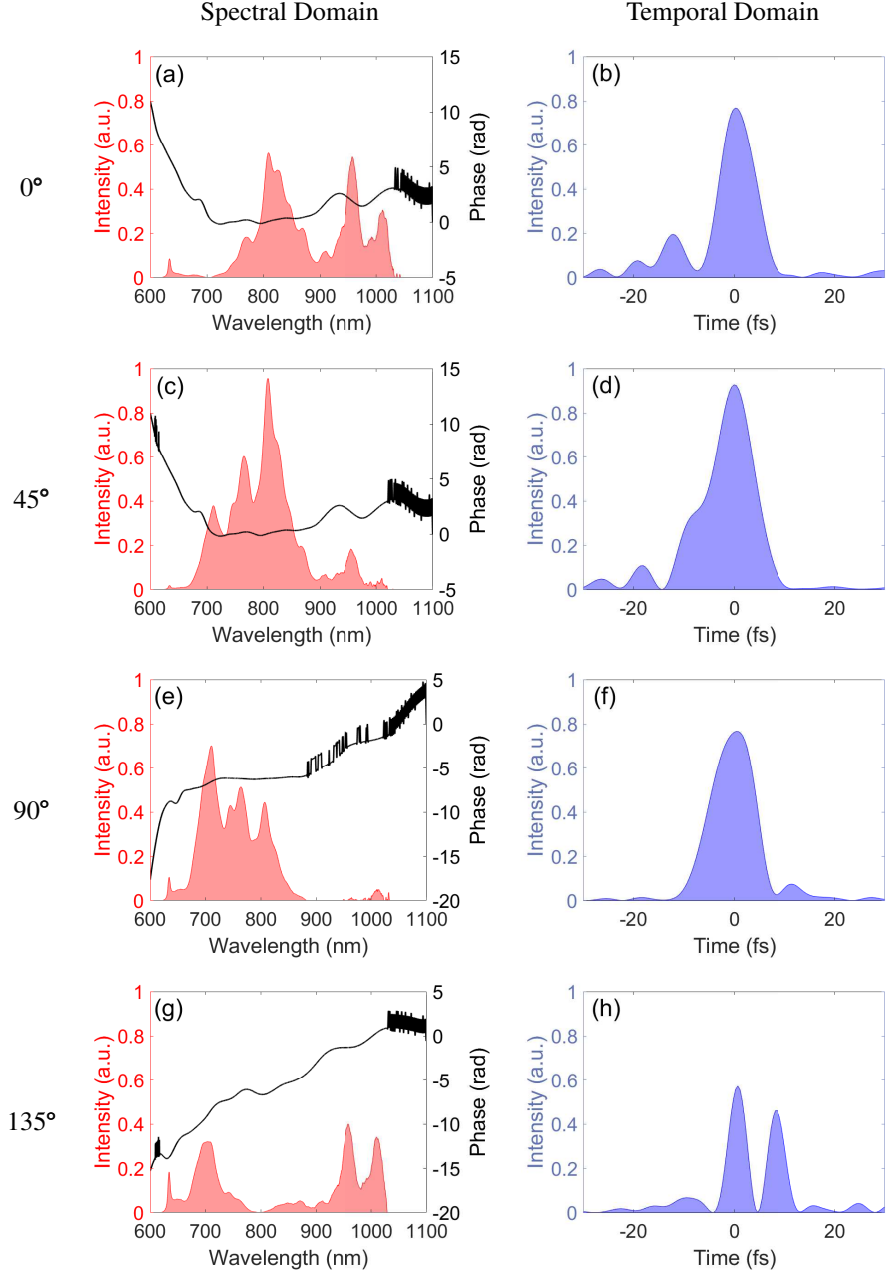

\centering
\begin{tabular}{ccc}
       & Spectral Domain & Temporal Domain \\
$0^{\pmb{\circ}}$  & \imagea & \imageb \\
$45^{\pmb{\circ}}$  & \imagec & \imaged \\
$90^{\pmb{\circ}}$  & \imagee & \imagef \\
$135^{\pmb{\circ}}$  & \imageg & \imageh \\
\end{tabular}
\caption{Retrieved spectral phases and temporal intensities for four projections of the polarization gate: (a-c-e-g) measured spectrum and retrieved spectral phase; (b-d-f-h) temporal intensity. These retrievals correspond to the material insertion that produces the most optimally compressed pulse.}
\label{fig:retrieved_intensity_and_phase}
\end{figure}

The retrieval algorithm extracts the spectral phase for each of the four linearly polarized projections of the electric field. Given the measured spectrum and the retrieved spectral phase, the pulse profile in the time domain can be computed using Fourier transform. Figure~\ref{fig:retrieved_intensity_and_phase} presents the intensity and phase in both the spectral and temporal domains for the four projections of the pulse. It should be noted that the retrieved spectral phase is obtained at the position indicated by the dashed line in the retrieved traces of Fig.~\ref{fig:dscan_measurements}, third column.
\par
As observed in Fig.~\ref{fig:retrieved_intensity_and_phase}, the pulse projection at $135^\circ$ exhibits a double-peaked structure, indicating that the pulse is linearly polarized at $45^\circ$. This result demonstrates the d-scan's sensitivity to temporal variations in the polarization state. The results obtained for the projections at $0^\circ$, $45^\circ$, and $90^\circ$ show a pulse intensity distribution with a single peak. This occurs because none of these projections are orthogonal to the pulse polarization at any time, resulting in a pulse envelope with a lower intensity but a shape similar to the input pulse.

\subsection{Reconstruction of the Time-Dependent Polarization State}

To reconstruct the time-dependent polarization state, only three of the previous measurements are needed. The projections at $0^\circ$ and $90^\circ$ are chosen as the orthogonal components, and the projection at $135^\circ$ is used to determine the parameters $\tau$ and $\varphi$ using the method described in Section~\ref{sec:reconstruction_polarization_state}.
\par
\begin{figure}[H]
    \centering
    \begin{subfigure}{0.47\textwidth}
        \includegraphics[width=\linewidth]{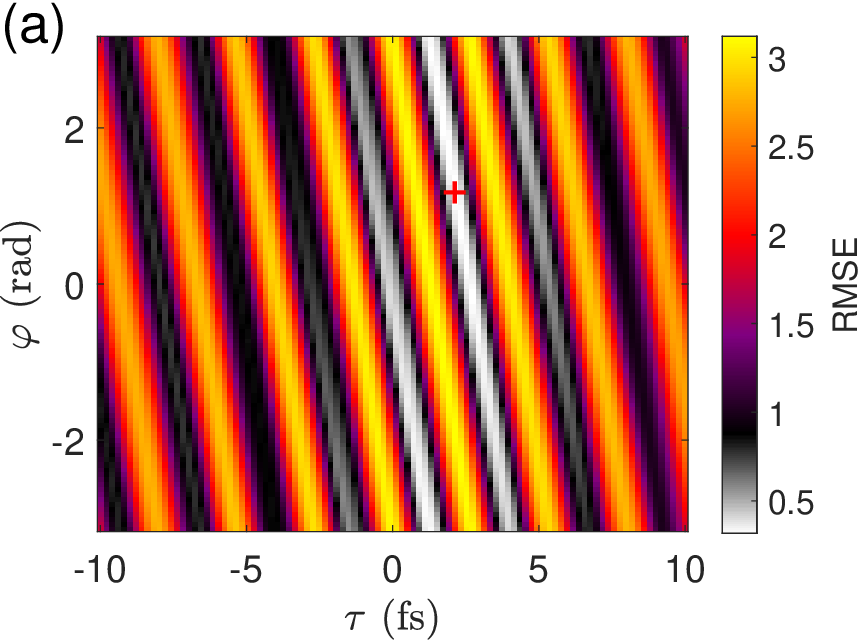}
    \end{subfigure}
    \hfill
    \begin{subfigure}{0.47\textwidth}
        \includegraphics[width=\linewidth]{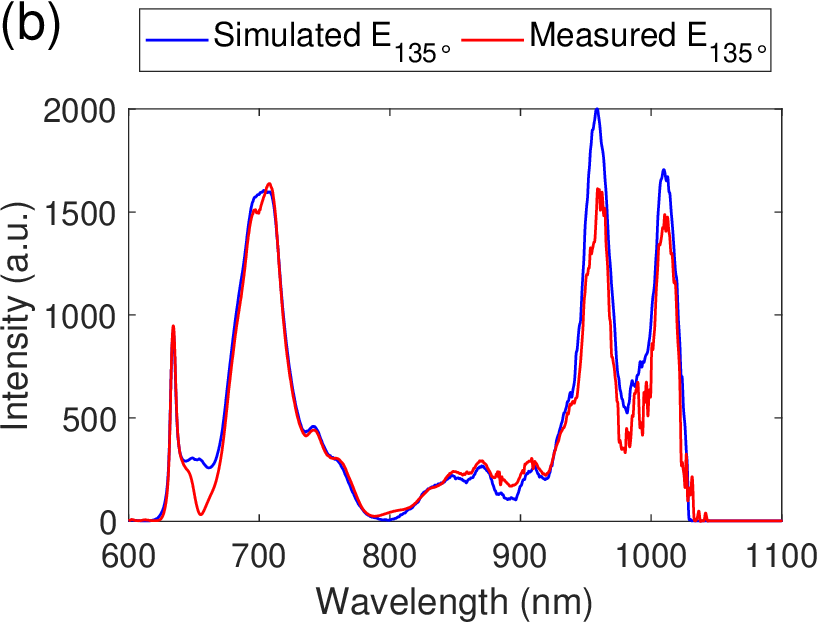}
    \end{subfigure}
    \caption{Optimization of delay ($\tau$) and relative phase difference ($\varphi$) values: (a) A colormap visualization of Root Mean Square Error (RMSE) values across a grid of $\tau$ and $\varphi$ combinations. The position of the minimum RMSE value is denoted by the red cross. (b) A comparative plot illustrating the optimal simulated spectra (based on the optimal $\tau$ and $\varphi$ values) and corresponding measured spectra.}
    \label{fig:optimization}
\end{figure}

Figure~\ref{fig:optimization}(a) displays the root mean square error (RMSE) calculation for all points of $\tau$ and $\varphi$ within the constructed grid. The minimum RMSE value (red cross) is attained for $\tau = \SI{2.12}{fs}$ and $\varphi = \SI{1.17}{rad}$, indicating the optimal values for reconstructing the polarization state.
\par
Figure~\ref{fig:optimization}(b) showcases the spectrum calculated using these parameters and compares  it with the spectrum obtained from the d-scan measurement. Although the two spectra do not match perfectly, there is a reasonably high degree of agreement.
\par
\begin{figure}[H]
    \centering
    \begin{subfigure}{0.47\textwidth}
        \includegraphics[width=\linewidth]{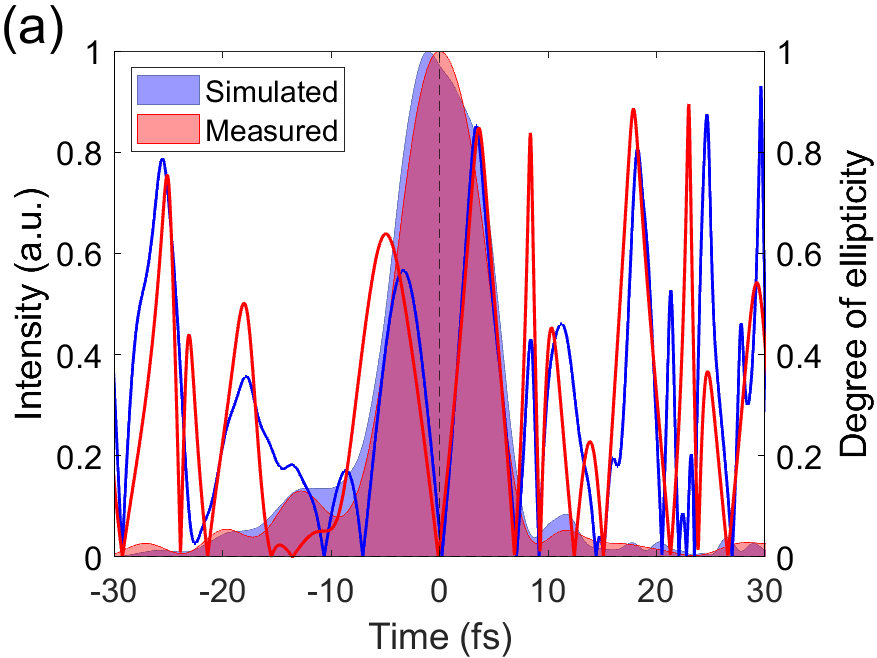}
    \end{subfigure}
    \hfill
    \begin{subfigure}{0.47\textwidth}
        \includegraphics[width=\linewidth]{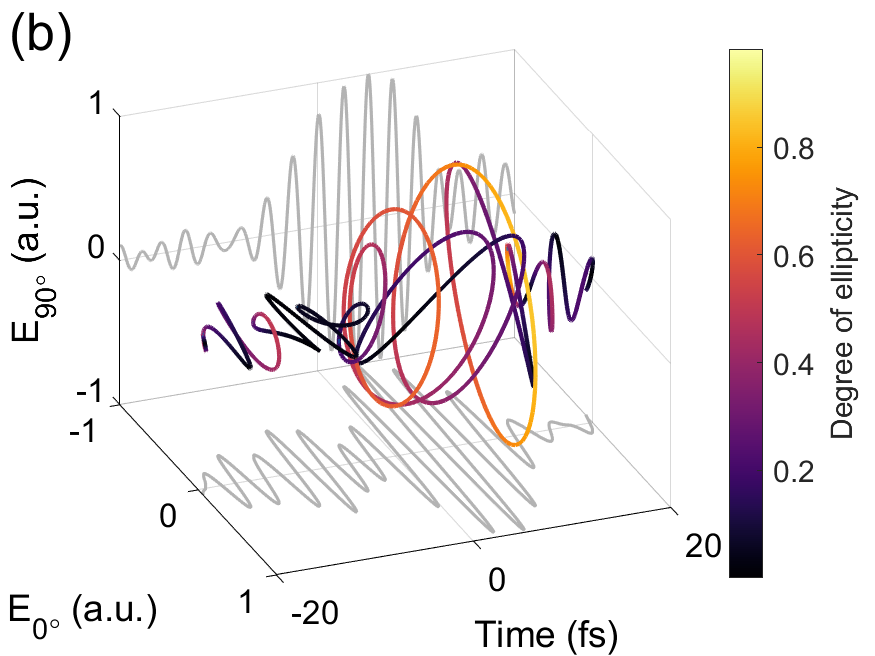}
    \end{subfigure}
    \caption{Reconstruction of the time-dependent polarization state: (a) The shaded blue and red areas represent the simulated and measured envelopes of the polarization gate, respectively. The blue and red lines represent the degree of ellipticity for the simulated and measured polarization gate, respectively. (b) The colored line represents the evolution of the electric field endpoint as a function of time, with the color indicating the degree of ellipticity of the polarization state. The lines on the bottom and in the background correspond to the horizontal and vertical projections of the electric field, respectively.}
\label{fig:polarization_gate_reconstruction}
\end{figure}

Figure~\ref{fig:polarization_gate_reconstruction} finally presents the reconstruction of the time-dependent polarization state from two different perspectives. In Fig.~\ref{fig:polarization_gate_reconstruction}(a), both the envelope of the measured polarization gate pulse and the measured ellipticity are depicted alongside the calculated envelope and ellipticity derived from the theoretical model for comparison. Within the main portion of the pulse, the two pulse shapes and ellipticity curves demonstrate a high degree of overlap and strong agreement. A more noticeable discrepancy between the measurement and the theoretical model emerges in areas where the pulse intensity is low. As reported in \cite{Tcherbakoff2003}, the emission efficiency of plateau harmonics in high-order harmonic generation decreases by $50\%$ when the degree of ellipticity of the fundamental driving field is more than $0.13$. This threshold can be used as a criterion for defining the width of the polarization gate where harmonics are efficiently generated. The measured envelope of the polarization gate yields a temporal width estimate of $\SI{1.36}{fs}$, which is approximately equivalent to one half cycle of the fundamental field. Additionally, Fig.~\ref{fig:polarization_gate_reconstruction}(b) provides a three-dimensional representation of the measured electric field. The line color indicates the ellipticity, which can be readily observed by examining the oscillation of the electric field. At the beginning (negative time), the pulse is circularly polarized; at the center, it is linearly polarized at approximately $45^\circ$; and at the end (positive time), it becomes circularly polarized with the opposite handedness.
\par

\section{Conclusion}
\label{sec:conclusion}

In this work, we proposed and implemented a novel approach for measuring time-dependent polarization states using the d-scan method. This strategy enabled the successful characterization of a polarization gate within the few-cycle regime. This technique involved measuring both the amplitude and phase for two orthogonal projections of the electric field, followed by determining the absolute phase difference and delay from the measured spectrum at an intermediate angle. Simulations validated the results, demonstrating an excellent agreement.
\par
Unlike other techniques, the d-scan is an in-line method, eliminating the need for interferometric precision. This attribute makes it more accessible and less susceptible to environmental interference. Additionally, the d-scan has established itself as a robust and stable method for measuring pulse parameters. Integrating the measurement of time-dependent polarization states into the d-scan technique will enhance its capabilities, giving more comprehensive insights into the temporal behaviour of ultrashort laser pulses.


\begin{backmatter}

\bmsection{Funding}
The authors acknowledge support from the Swedish Research Council (2013-8185, 2021-04691), the European Research Council (advanced grant QPAP, 884900), the European Innovation Council (101058075 - SISHOT), European Union's Horizon 2020 research and innovation programme (871124 Laserlab-Europe) and the Knut and Alice Wallenberg Foundation. AL is partly supported by the Wallenberg Center for Quantum Technology (WACQT) funded by the Knut and Alice Wallenberg Foundation

\bmsection{Acknowledgments}
The authors acknowledge fruitful discussion with O.~Cohen, G.~I.~Haham, I.~J.~Sola, B.~Alonso, M.~Miranda, R.~Romero, H.~Crespo and P.~T.~Guerreiro. 

\bmsection{Disclosures}
The authors declare no conflicts of interest.

\bmsection{Data availability}
Data underlying the results presented in this paper are not publicly available at this time but may be obtained from the authors upon reasonable request.

\end{backmatter}

\bibliography{references}

\end{document}